\documentclass[a4paper,fleqn,usenatbib]{mnras}

\usepackage{newtxtext,newtxmath}
\usepackage{hyperref}

\usepackage[T1]{fontenc}
\usepackage{ae,aecompl}

\usepackage{graphicx}	
\usepackage{amsmath}	
\usepackage{amssymb}	
\usepackage{physics}	

\title[HD$\,$169142 in the eyes of ZIMPOL/SPHERE]{HD$\,$169142 in the eyes of ZIMPOL/SPHERE}

\author[G. H.-M. Bertrang et al.]{G. H.-M. Bertrang,$^{1,2,3}$\thanks{E-mail: bertrang@das.uchile.cl}
H. Avenhaus,$^{1,3,4}$
S. Casassus,$^{1,3}$
M. Montesinos,$^{1,3,5}$
\newauthor F. Kirchschlager,$^{6}$
S. Perez,$^{1,3}$
L. Cieza,$^{2,3}$
S. Wolf$^{7}$
\\
$^{1}$Universidad de Chile, Departamento de Astronom\'ia, Casilla 36-D, Santiago, Chile\\
$^{2}$Universidad Diego Portales, Facultad de Ingenier\'ia, Av. Ej\'ercito 441, Santiago, Chile\\
$^{3}$Millennium Nucleus Protoplanetary discs in ALMA Early Science, Universidad de Chile, Casilla 36-D, Santiago, Chile\\
$^{4}$ETH Zurich, Institute for Particle Physics and Astrophysics, Wolfgang-PauliStrasse 27, CH-8093, Zurich, Switzerland\\
$^{5}$Instituto de Astrof\'isica, Facultad de F\'isica, Pontificia Universidad Cat\'olica de Chile, 782-0436 Santiago, Chile\\
$^{6}$Department of Physics and Astronomy, University College London, Gower Street, London WC1E 6BT, UK\\
$^{7}$Kiel University, Institute of Theoretical Physics and Astrophysics, Leibnizstr. 15, 24118 Kiel, Germany
}

\date{Accepted XXX. Received YYY; in original form ZZZ}

\pubyear{2017}

\AtBeginShipout{%
  \ifnum\value{page}>1 %
    \typeout{* Additional boxing of page `\thepage'}%
    \setbox\AtBeginShipoutBox=\hbox{\copy\AtBeginShipoutBox}%
  \fi
}

\begin{document}
\label{firstpage}
\pagerange{\pageref{firstpage}--\pageref{lastpage}}
\maketitle

\begin{abstract}
We present new data of the protoplanetary disc surrounding the Herbig Ae/Be star HD$\,$169142 obtained in the very broad-band (VBB) with the Zurich imaging polarimeter (ZIMPOL), a subsystem of the Spectro-Polarimetric High-contrast Exoplanet REsearch instrument (SPHERE) at the Very Large Telescope (VLT). Our Polarimetric Differential Imaging (PDI) observations probe the disc as close as $0\farcs03$ ($3.5$au) to the star and are able to trace the disc out to $\sim$1\farcs08 ($\sim 126$ au). We find an inner hole, a bright ring bearing substructures  around $0\farcs18$ ($21$au), and an elliptically shaped gap stretching from $0\farcs25$ to $0\farcs47$ ($29-55$au). Outside of $0\farcs47$, the surface brightness drops off, discontinued only by a narrow annular brightness minimum at $\sim0\farcs63-0\farcs74$ ($74-87$au). These observations confirm features found in less-well resolved data as well as reveal yet undetected indications for planet-disc interactions, such as small-scale  structures, star-disk offsets, and potentially moving shadows.
\end{abstract}




\section{Introduction}
\begin{table*} 
\caption{Observation summary.}
\label{obstab}
\begin{center}
\begin{tabular}{l c c c c c c}\\\hline\hline
Night & Filter~$1$ & Filter~$2$ & Read-out  & Polarization  & DIT &Total time \\
          &                  &                  &  mode                & mode   & [s] & [s]                \\\hline

July~$8, 2015$ & R (I) * & I (R) * & slowPol & P$1$ & 10 & $2640\,$\\
July~$8, 2015$ & R (I) * & I (R) * & fastPol & P$1$ & 6 & $1728\,$\\
July~$9, 2015$ & VBB * & VBB * & fastPol & P$2$ & 10 & $3360\,$\\\hline\hline
\multicolumn{7}{l}{* The R and I band filters were interchanged during the observation, but the data was reduced together.}
\end{tabular}
\end{center}
\end{table*}

Circumstellar discs around young stars are the key to understand star and planet formation \citep[e.g.,][]{2011ARA&A..49...67W}. However, due to new high-contrast and high-sensitivity instruments, the study of star and planet formation is undergoing a rapid change with new spectacular observations of structures in protoplanetary discs such as gaps \citep[e.g.,][]{2013ApJ...766L...2Q} and spiral structures that appear in the scattered light \citep[e.g.,][]{2014ApJ...781...87A, 2015A&A...578L...6B}, where observations at long wavelengths reveal signs of dust traps \citep[e.g.,][]{2015ApJ...813...76M}.
These recent observations glimpse at a sophisticated picture of disc evolution and dispersal, which in turn regulates the origin of planetary systems. 

The protoplanetary disc around the young ($6_{-3}^{+6}\,$Myr) and isolated Herbig Ae/Be star HD$\,$169142 (M$_{\ast} = 2\,$M$_{\odot}$, T$_{\rm eff} = 8100\,$K) is a well-studied object  at a distance of only $117\pm 4\,$pc \citep{2007ApJ...665.1391G, 1994A&AS..104..315T, 2006ApJ...653..657M, 2016A&A...595A...2G}. Previous estimates by \cite{1999AJ....117..354D} reported instead a distance of $145\,$pc. In the rest of the paper we will use the newest distance estimate adjusting all the relevant parameters. CO observations reveal an almost face-on orientation of the disc with an inclination of $13\pm1^{\circ}$ \citep{2006AJ....131.2290R}. Within recent years, observations from the near-infrared to the mm-wavelength range revealed an inner cavity with a radius of $\sim 16\,$au and an annular gap present at $\sim 32-56\,$au \citep{2013ApJ...766L...2Q, 2015PASJ...67...83M, 2017A&A...600A..72F}.  \cite{2014ApJ...791L..36O} found indications for a residual disc in $\sim 0.16 - 0.48\,$au distance from the central star, according to which the central cavity described by \cite{2013ApJ...766L...2Q} is, in fact, the inner most gap of HD$\,$169142. A third, narrow gap at $\sim 85\,$au was detected in recent VLA observations at $7\,$mm and $9\,$mm and is associated with the CO~snowline  \citep{macias2017}. 

A number of mechanisms, such as magneto-rotational instabilities creating dead-zones \citep{2015A&A...574A..68F}, giant planets carving out dust, photo-evaporation or dust grain growth  \citep[][and references therein]{2014prpl.conf..497E} have been proposed to explain cavities and gaps in circumstellar discs. Yet, there are observational constraints which favour the creation of most cavities found in transition discs due to dynamical interactions with substellar or planetary companions \citep{2011ApJ...732...42A, 2014prpl.conf..497E}. In three independent observations of HD$\,$169142, a substellar or planetary companion candidate was detected at $0\farcs105\pm 5\,$mas (P.A.$\,=4^{\circ}$) applying PDI and coronagraphic imaging, respectively \citep[][the finding by \cite{2014ApJ...792L..23R} is slightly offset but consistent]{2014ApJ...792L..22B, 2018MNRAS.473.1774L}. Additionally, \cite{2017A&A...600A..72F} discuss different scenarios for a system of multiple giant planets as most likely for HD$\,$169142. 

In this work, we present our new SPHERE/ZIMPOL data on HD$\,$169142 which confirm the previously found gap structure as well as reveal yet undetected indications for planet-disc interactions in the form of sub-structures and offsets in the dust distribution of the inner part of the disc.


\section{Observations and Data Reduction}

\begin{figure*}
\centering
\includegraphics[width=\linewidth]{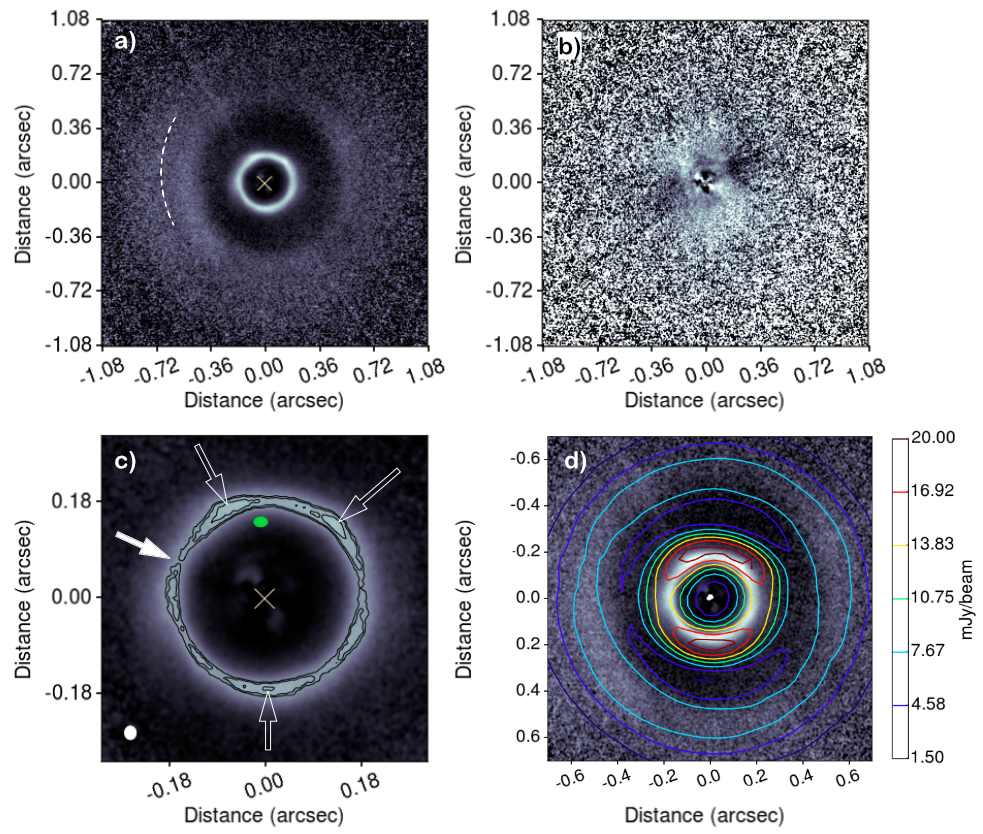}
\caption{ZIMPOL/SPHERE PDI observations of HD$\,$169142 in VBB. Final $Q_\varphi$ {\bf (a)} and $U_\varphi$ {\bf (b)}. The dashed line in (a) indicates a $2\sigma$~detection of a shallow and narrow gap which was first detected at millimeter wavelengths \citep{macias2017}. In {\bf (c)}, we show a zoom onto the ring of $Q_\varphi$  where contour lines indicate $20\%, 22.5\%$, and $25\%$ levels of the peak flux. The filled arrow marks the dip discussed in Section~\ref{res}, and the hollow arrows mark the peaks discussed in Section~\ref{discussion}. The stellar position is marked with a cross, the green ellipse indicates the position of the protoplanet candidate \citep{2014ApJ...792L..22B, 2018MNRAS.473.1774L}, and the white ellipse displays the spatial resolution of this observation. {\bf (d)} Superposition of $Q_\varphi$ with contour lines of ALMA band~$7$ $(880\,\mu{\rm m}$,  beam size: $0\farcs14)$ continuum data (archival data from program $2012.1.00799.$S). 
All maps are displayed in $r^2$ scaling, North is up and East is left.}
\label{pic:obs1}
\end{figure*}

\begin{figure*}
\centering
\includegraphics[width=\linewidth]{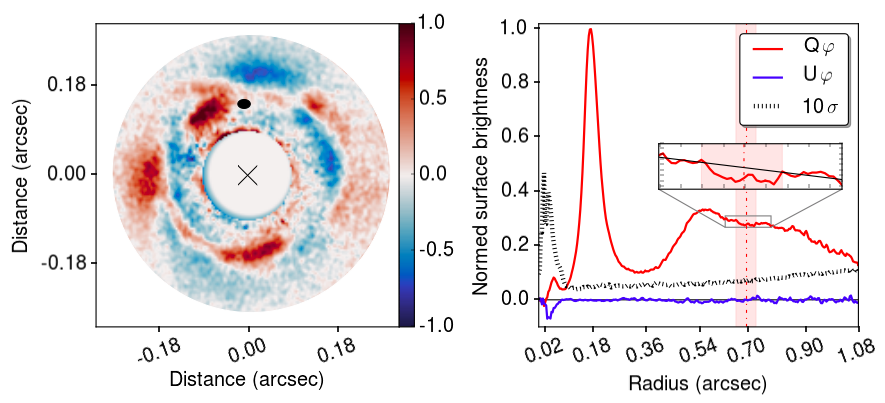}
\caption{ {\bf Left:} The bright inner ring in $Q_\varphi$ subtracted by the azimuthal average revealing the clear sub-structures in this part of the disc. The azimuthal average in this panel is derived relative to the offset center of the inner ring (Section~\ref{res}). Cross and black ellipse mark the position of the central star, resp. the protoplanet candidate \citep{2014ApJ...792L..22B, 2018MNRAS.473.1774L}. The map is displayed in $r^2$ scaling, North is up and East is left. {\bf Right:} Azimuthally averaged radial surface brightness profile of $Q_\varphi$ (red solid line) and $U_\varphi$ (blue solid line), as well as $10\sigma$ level (black dashed line). The azimuthal average in this panel is derived relative to the central star. The center of the outer dip in the surface brightness distribution is marked by red dashed-dotted line, the red shaded region marks its extend. The brightness peak at its center results from the two instrumentally induced spikes visible in Figure~1(a).}
\label{pic:obs2}
\end{figure*}

\begin{figure*}
\centering
\includegraphics[width=\linewidth]{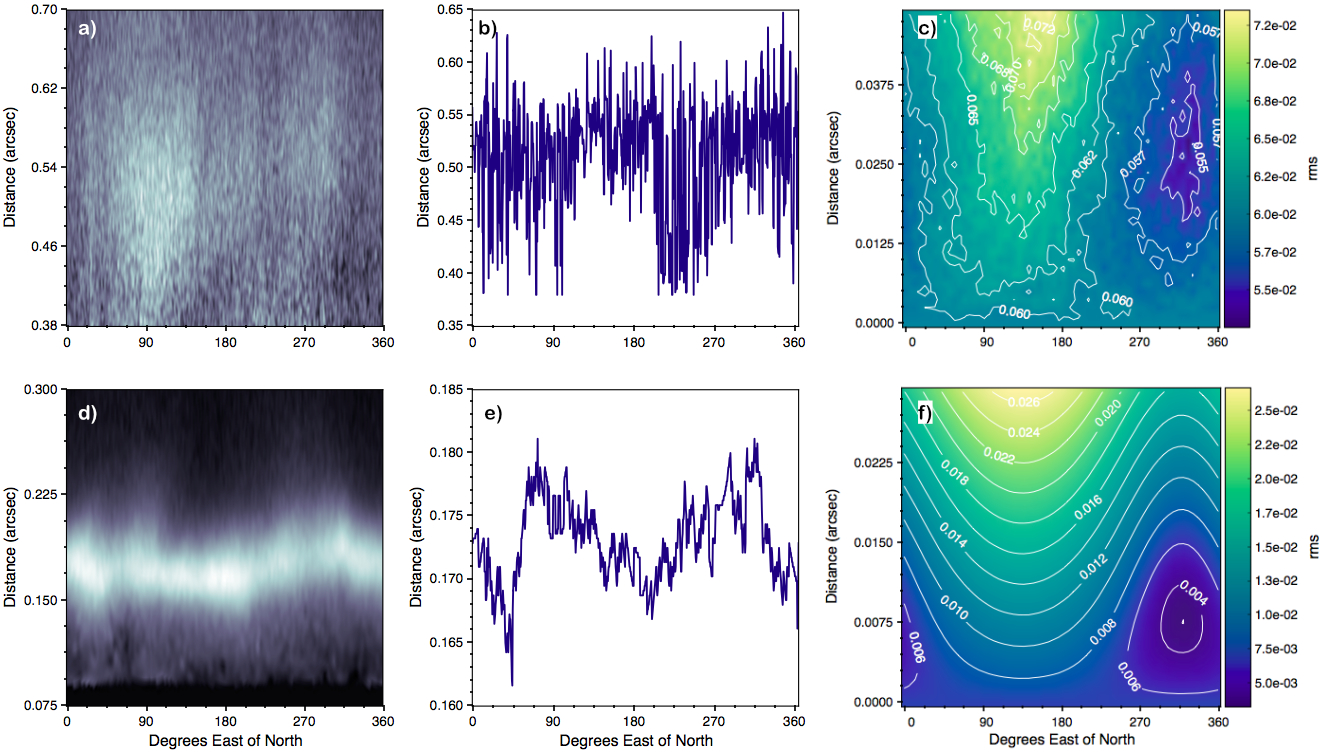}
\caption{Analysis of the outer disc ({\bf upper row}) and inner ring ({\bf lower row}) in polar coordinates: {\bf (a) \& (d):} Maps of $Q_{\varphi}$, scaled by $r^2$; {\bf (b) \& (e):} radial brightness maxima.  {\bf(c) \& (f):} Grid search for the minimum rms of the difference between the inner ring, resp. outer disc with an offset in $(r,\phi)$ direction and an inclined circle. The assumed inclination is $13^{\circ}$, the position angle is P.A.$\,=5^{\circ}$ \citep{2006AJ....131.2290R}. }
\label{pic:polarplot}
\end{figure*}

The data were taken during the nights of July 8 and 9, 2015, at the ESO Very Large Telescope situated on Cerro Paranal, Atacama desert, Chile. The instrument used was the Zurich Imaging Polarimeter (ZIMPOL) sub-instrument of SPHERE \citep[Spectro-Polarimetric High-contrast Exoplanet REsearch;][]{2008SPIE.7014E..18B}. During the first night, data were taken in the P1 mode of ZIMPOL (field rotates), while during the second night, the P2 mode was used (pupil stabilized, no field rotation, for more details consult the SPHERE user manual). The filters used were the R and I band filters in the two arms of ZIMPOL for the P1 mode observations (both filters can be used simultaneously due to the construction of ZIMPOL), and the VBB/RI filter (very broad band encompassing both the R and I band, central wavelength: $0.73\,\mu$m, see SPHERE manual) in both arms for the P2 mode observations. For the P1 mode observations, both SlowPol and FastPol observations were performed, while the P2/VBB observations were only taken in FastPol mode to allow for optimal contrast. Table 1 summarizes the observations.

The data is pre-processed in the standard way (bias frame subtraction, flat fielding, bad pixel correction). Two bad lines on the detector appearing in the FastPol mode have been mapped out manually. On these corrected frames, the stellar position is then determined by fitting a 2-dimensional Gaussian. Because ZIMPOL pixels are unusual in a way that they cover a non-square area of the sky due to the construction of ZIMPOL which has each second row covered and the pixels thus each correspond to an on-sky area of $\sim7\farcs2 \times 3\farcs6$, we use a skewed 2D Gaussian (twice as wide in one direction). The data is then re-mapped onto a square pixel grid ($3\farcs6 \times 3\farcs6$) and centered on the pre-determined stellar position. Instrumental effects (difference between ordinary and extraordinary beam) are corrected as described in the appendix of \cite{2014ApJ...781...87A}, using inner and outer correction radii of $0\farcs25$ and $0\farcs35$, respectively. After this, an astrometric calibration is performed, accounting for the slight differences in pixel scale along the detector axes and the True North offset.

Finally, the Stokes parameters $Q$ and $U$, representing linearly polarised intensities, are calculated using the double ratio method \citep[see][and references therein]{2014ApJ...781...87A}. The local Stokes vectors, now called $Q_{\varphi}$ and $U_{\varphi}$ by most authors \citep[e.g., ][]{2015A&A...578L...6B} are calculated as:

\begin{equation}
\begin{split}
Q_{\varphi}&=+Q\cos(2\varphi)+U\sin(2\varphi),\\
U_{\varphi}&=-Q\sin(2\varphi)+U\cos(2\varphi),\\
\varphi&=\arctan\frac{x-x_0}{y-y_0}+\theta
\end{split}
\end{equation}

\noindent  where $(x,y)$ gives the pixel coordinate, $(x_0, y_0)$ is the center coordinate, and $\varphi$ represents the azimuth angle. Here, $\theta$ is used to correct for the fine-alignment of the half-wave plate (HWP) rotation and is determined from the data by assuming that $U_{\varphi}$ should on average be zero (note that while $U_{\varphi}$ does not have to be zero everywhere due to e.g. multiple scattering, e.g. \cite{2015A&A...582L...7C}, in a symmetric disc it will be zero on average for reasons of symmetry).


\section{Results and Analysis}\label{res}

Figure~\ref{pic:obs1} shows the final $Q_{\varphi}$ and  $U_{\varphi}$ images. A map of $Q_{\varphi}$ in polar coordinates is shown in Figure~\ref{pic:polarplot}. The data are not flux calibrated, thus, we are limited to an analysis of the relative surface brightness distribution.

We resolve the protoplanetary disc around HD$\,$169142 on a scale of $0\farcs0230\,\times\,0\farcs0233$ ($2.69\,\times\,2.72\,$au). By this, we reach an unprecedented spatial resolution for this disc. Our observation probe it as close as $0\farcs03$ ($\sim3.5\,$au) to the star,  though we do experience some increased noise inside of $\sim0\farcs1$, and out from 
$\sim1\farcs08$ ($\sim 126\,$au). We detect clear structures in the scattered light image of the local Stokes vector $Q_{\varphi}$ consistent with previous studies \citep[e.g., ][]{2013ApJ...766L...2Q, 2015PASJ...67...83M, 2017ApJ...838...20M, 2017A&A...600A..72F, 2017ApJ...850...52P}. The $U_{\varphi}$  image does not contain any significant signal as it is expected from a disc with low inclination \citep[e.g.,][]{2013ApJ...766L...2Q, 2015A&A...582L...7C}. 

Close to the star (Figure~\ref{pic:obs1},~\ref{pic:obs2},~\ref{pic:polarplot}) the flux is low and increases until its peak around $0\farcs173$ ($\sim20\,$au). The peak position of the bright ring varies with azimuthal angle (see Figure~\ref{pic:polarplot}). The narrow ring ($\Delta\sim0\farcs059\,; \text{corresponding to} \sim7\,$au) is surrounded by an elliptically shaped gap  (P.A.$\,=5^{\circ}, e \approx 0.5$) which stretches from $\sim0\farcs25$ ($\sim 29\,$au) to $\sim0\farcs38$ ($\sim45\,$au) which has been described so far, based on lower resolution observations, as annular \citep{2013ApJ...766L...2Q}. Outside of $\sim0\farcs47$  ($\sim55\,$au) the surface brightness of the disc drops off, discontinued by a narrow annular brightness minimum at $\sim0\farcs63-0\farcs74$ ($\sim74-87\,$au)  detected at a  $2\sigma$ level.  This shallow and narrow ring coincides with a gap found at $7\,$mm and $9\,$mm which is associated with the CO~snowline \citep{macias2017}. 

The bright inner ring is spatially resolved and appears structured and circular (Figure~\ref{pic:obs1}), which is in contrast to the elliptically shaped gap surrounding it. Its width ($\sim 7\,$au) is similar to that reported by \cite{macias2017} at $7\,$mm ($8\,$au). We find a brightness dip at  P.A.$\sim 50^{\circ}$. If this is the same dip reported by \cite{2013ApJ...766L...2Q} at $\sim 80^{\circ}$, then the orbital velocity of this feature is $\sim 16\,$km/s. A dip with this orbital velocity can be produced by a large structure in Keplerian motion around the central star. An object such as a jupiter-mass planet or a brown dwarf surrounded by a small accretion disk at $\sim 6\,$au from the central star is a possible candidate for this speculative scenario \citep[][and references therein]{2016yCat..35980043C}. Such an object would explain the formation of the inner cavity in HD$\,$169142 and contribute to the residual dust close to the central star which has not yet been spatially resolved. Further, we detect a weak signal in East-West direction inside the gap which probably is a convolution effect as our radiative transfer simulations show (see Section~\ref{discussion}). This relatively narrow ring is more extended in the ALMA observation at $880\mu$m, however, the surrounding gap remains clearly defined (Figure~\ref{pic:obs1}$\,$d).

Figure~\ref{pic:polarplot} shows the radial maximum of the surface brightness alternating with azimuth angle. This is found for both the bright inner ring as well as the inner rim of the outer disc. In the following, we analyse this behaviour in more detail. For this, we presume a radially symmetric geometry of the disc and fit a ring to the observed radial maximum brightness position for (i) the bright inner ring and (ii) the inner rim of the outer disc. This ring is inclined by $13^{\circ}$  with a position angle of $5^{\circ}$ in consistence with the disc around HD$\,169142$ \citep{2006AJ....131.2290R} and centred on the star. We find that such a ring fits the observed structure in both cases only if it is off-centred with respect to the star. In order to determine a robust offset in radius and azimuth, we run a grid search with the aim to minimise the rms of the fit.  For this, we determine the stellar position in our data with an accuracy of $1.8\,$mas and resolve the grid search in azimuthal direction by $5.625^{\circ}$ and in radial direction by $0.75\,$mas (bright inner ring), resp. $1.25\,$mas (outer disk). Figure~\ref{pic:polarplot} displays the rms as a function of radial distance and azimuth of the center of the fit. We find that the offset for both the bright inner ring and the inner rim of the outer disc is directed towards North with a significant difference in the radial offsets. The rms minimises for the center of the inner ring (outer disc) at P.A.$=315^{\circ}$  ($320.6^{\circ}$) at a radial distance of $\sim7.5\,$mas ($\sim 287.5\,$mas) from the star - an indication for planet-disc interactions.

\begin{figure*}
\centering
\includegraphics[width=0.7\linewidth]{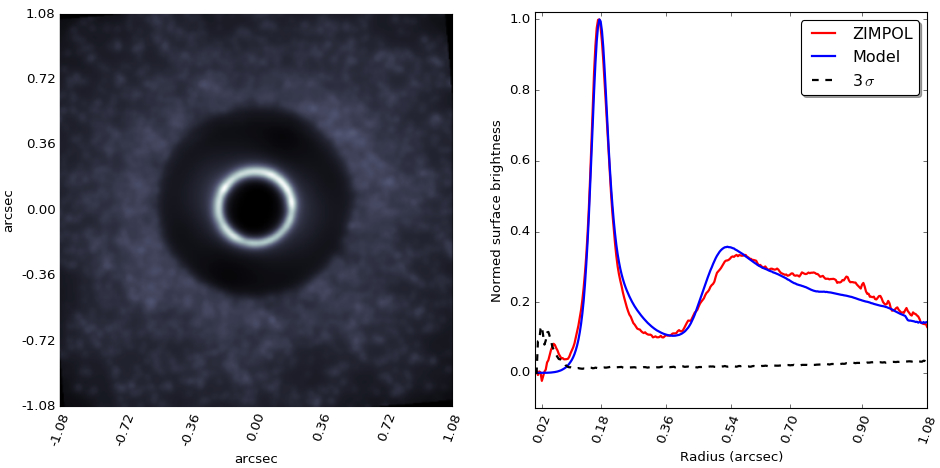}
\includegraphics[width=0.7\linewidth]{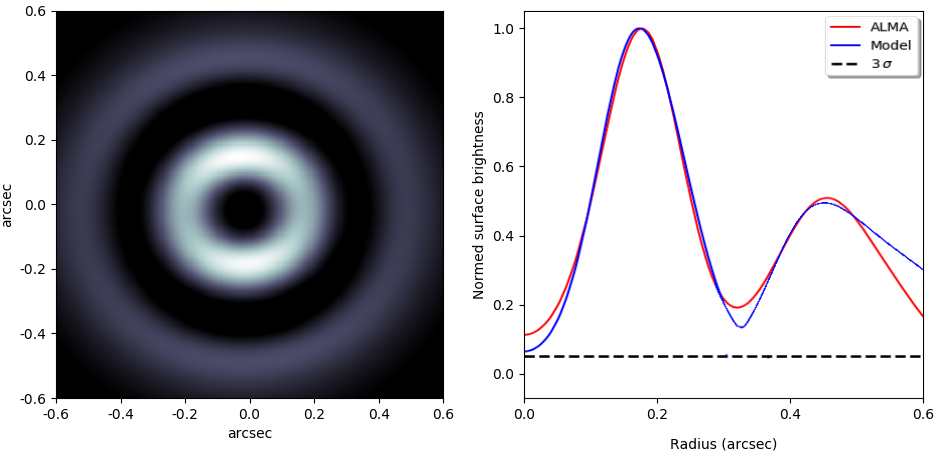}
\caption{Results of our radiative transfer simulations (see section~\ref{rt}): {\it Top row:} $Q_{\varphi}$ map convolved with the ZIMPOL beam ({\it left}). {\it Right:} Azimuthally averaged radial profiles comparing the ZIMPOL data with our RT model. {\it Bottom row:} Intensity map convolved with the beam of the ALMA band~$7$ continuum data (from program $2012.1.00799.$S; {\it left}). {\it Right:} Azimuthally averaged radial profiles comparing the ALMA data with our RT model.   The azimuthal averages in this figure are derived relative to the central star.}
\label{pic:sims}
\end{figure*}
 
\section{Radiative transfer model} \label{rt}

\begin{figure}
\centering
\includegraphics[width=\linewidth]{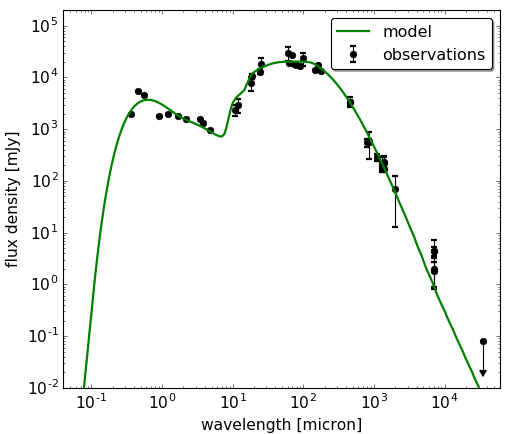}
\caption{Observed (points) and model (line; see section~\ref{rt}) spectral energy distributions of HD$\,$169142. The photometric data are taken from  \citet{2006MNRAS.365.1283D}, \citet{2006AJ....131.2290R}, \citet{2012ApJ...752..143H}, \citet{2010A&A...518L.124M}, \citet{2011ApJ...727...26S}, \citet{2013MNRAS.430.2534D}, \citet{2014ApJ...791L..36O}, \citet{macias2017}, and \citet{2017A&A...600A..72F}. The error of the photometric data is given in $3\sigma$.}
\label{pic:sed}
\end{figure}

HD$\,$169142 has been observed, in part spatially resolved \citep[e.g.,][ this work]{2013ApJ...766L...2Q, 2014ApJ...792L..23R, macias2017}, across a wide spectrum from the ultra-violet to the radio wavelength range; Figure~\ref{pic:sed} shows the spatially unresolved photometric measurements.  In order to understand the underlying physical scenario, several previous studies present models based on a variety of numerical approaches  \citep[e.g.,][]{2005ApJ...630L.185C, 2006ApJ...638..314D, 2010A&A...518L.124M, 2013A&A...555A..64M, 2014ApJ...791L..36O, 2017A&A...600A..72F}. Motivated by the newly revealed details of HD$\,$169142 presented in this work, we aim at refining those previous models, especially \cite{2014ApJ...791L..36O} as it is one of the most comprehensive ones, by fitting the here presented spatially resolved data as obtained with SPHERE/ZIMPOL and ALMA, its radial surface brightness profile, as well as the spectral energy distribution of HD$\,$169142. For this purpose, we apply the 3D~radiative transfer code {\it Mol3D} \citep[][]{2015A&A...579A.105O}. Based on the previous models, we adapt a radial symmetric density distribution of an accretion disc \citep{1973A&A....24..337S}, given here in cylindrical coordinates:

\begin{equation}
\rho\left(r_z, z\right) = \rho_0 \left(\frac{r_0}{r_z}\right)^{\alpha}\exp\left(-\frac{1}{2}\left[\frac{z}{h\left(r_z\right)}\right]^2\right).
\end{equation}

The density parameter $\rho_0$ is determined by the total dust mass of the disc. The reference radius $r_0$ is set to $100\,$au, and the vertical scale height of the disc $h\left(r_z\right)$ is defined by

\begin{equation}
h\left(r_z\right) = h\left(r_0\right) \left(\frac{r_z}{r_0}\right)^{\beta}.
\end{equation}

The parameters $\alpha$ and $\beta$ define the radial density profile as well as the flaring of the disc and $h\left(r_0\right)$ is fixed to $10\,$au.  
The distribution of dust grain sizes, $a$, is described by a power law \citep[][]{1969JGR....74.2531D, 1977ApJ...217..425M}

\begin{equation}
\dv{N}{a} \sim a^{-3.5}.
\end{equation}

Motivated by \cite{2014ApJ...791L..36O}, we introduce two dust grain populations, small grains $(0.005-10\mu$m$)$ and large grains $(5-1000\mu$m$)$. The large grain population is settled around the mid plane with a radially increasing scale height for settling, $(\zeta\cdot r_z)$, where $r_z$ is the radius of the disc and $\zeta$ a free fitting parameter. If $\zeta=\infty$ the disk is in vertical direction fully filled with large grains.  In the ALMA band~7 data, the bright inner ring appears significantly wider compared to the ZIMPOL data  (see Figure~\ref{pic:obs1}). To take that into account, the radial extend of the large grain population in that region is fitted independently from that of the small grain population. Since HD$\,$169142 is a very well studied object, most of the parameters of the model are determined by previous observations or constrained pretty well and only need small adjustments. We adopted the literature model and manually optimised the parameters, especially the flaring parameters, to fit the SED as well as the spatially resolved data obtained with ZIMPOL and ALMA at the same time. The parameters of our best fit are listed in Table~\ref{paratab}  (see also Figures~\ref{pic:sims}~and~\ref{pic:sed}). Please note that our best fit model under\-estimates some of the photometric measurements at $7\,$mm. This is tolerated in order to simultaneously fit the bright inner ring and the sub-mm fluxes. In order to compare our simulations directly with the observations, we convolve the simulated Stokes parameters $Q$ and $U$ with the PSF of this data set and compute the displayed local Stokes vectors $Q_{\phi}$ subsequently. In doing so, we ensure to include the cancellation effect of polarimetry into our model. We find that a self-shadowed disc separated into three distinct parts can already explain some of the features found in HD$\,$169142 (Figure~\ref{pic:obs1},~\ref{pic:obs2}). Please note that the signal detected inside the gap in the ZIMPOL observations can be explained simply by an empty gap and the convolution with the PSF. However, this radially symmetric and parametric model is not able to explain the appearance of ring and elliptical structures at the same time, neither the offsets of several disc parts relative to the central star (see Section~\ref{res}), nor the sub-structures in the bright inner ring resolved in this observation for the first time. Such features hint towards local perturbations, such as caused by planet-disc interaction. To investigate this in more detail, we apply hydrodynamical simulations (see Section~\ref{hd}). 

\begin{table*} 
\caption{Model parameters for HD$\,$169142 and its disc. Note that distance-dependent values have been adjusted to the new distance estimate from \protect\cite{2016A&A...595A...2G}.}
\label{paratab}
\begin{center}
\begin{tabular}{l c c c}\\\hline\hline
Parameter & Value & Notes$^{*}$ & Ref.\\\hline
	& Stellar parameters & & \\\hline
Distance (pc)			& 117  & Adopted & 1\\
Effective temperature (K)	& 8100  & Adopted & 2\\
Radius (R$_{\odot}$)		& 1.76 & Adopted & 2 \\
Mass (M$_{\odot}$)		& 2.0  & Adopted & 2 \\\hline
	& disc parameters &   & \\\hline
Dust mass (M$_{\odot}$)	    & 1$\cross$10$^{-4}$  & \multicolumn{2}{l}{Fitted [10$^{-4}$, 10$^{-3}, 10^{-4}$]} \\
Grain size distr. 1 ($\mu$m) & 0.005-10   & Adopted/Refined & 3, 5\\
Grain size distr. 2 ($\mu$m) & 5-1000   & Adopted/Refined & 3, 5\\
Grain size distr. exponent	   & -3.5  & Adopted & 3\\  
Inclination ($^{\circ}$)	   & 13  & Adopted & 4 \\
Position angle  ($^{\circ}$)	   & 5  & Adopted &  4 \\\hline 
	& Inner most disc &  & \\\hline
r$_{\rm in}$ (au)     & 0.16   & Adopted/Refined & 5\\
r$_{\rm out}$ (au)  & 0.18   & Adopted/Refined & 5 \\
$\alpha$ 	      & 3.0  &  \multicolumn{2}{l}{Fitted [1.0, 4.0, $10^{-3}$]} \\
Flaring $\beta$ 	      & 1.125  &  \multicolumn{2}{l}{Fitted [1.0, 3.0, $10^{-3}$]} \\
$\zeta$                    & $\infty^{**}$  &  \multicolumn{2}{l}{Fitted [0.0, 2.0, $10^{-2}$]} \\\hline 
\hline
   	& Bright inner ring & &  \\\hline    
r$_{\rm in}$ (au) 	& 23.0  &  Adopted/Refined & 5\\
Grain size distr. 1: r$_{\rm out}$ (au) 	&  25.0  &  Adopted/Refined & 5\\
Grain size distr. 2: r$_{\rm out}$ (au) 	&  40.0  &  Adopted/Refined & 5\\
$\alpha$ 	      & 3.15  &  \multicolumn{2}{l}{Fitted [1.0, 4.0, $10^{-3}$]} \\
Flaring $\beta$ 		& 1.015 & \multicolumn{2}{l}{Fitted [1.0, 3.0, $10^{-3}$]} \\
$\zeta$                      & 0.1  &  \multicolumn{2}{l}{Fitted [0.0, 2.0, $10^{-2}$]} \\\hline 
   	& Outer disc &  & \\\hline
r$_{\rm in}$ (au) 	& 65.0   & Adopted/Refined & 5\\
r$_{\rm out}$ (au) 	& 194   & Adopted/Refined & 5\\
$\alpha$ 	      & 2.505  &\multicolumn{2}{l}{Fitted [1.0, 4.0, $10^{-3}$]} \\
Flaring $\beta$ 		&  1.005  & \multicolumn{2}{l}{Fitted [1.0, 3.0, $10^{-3}$]} \\
$\zeta$                      & 0.1  &  \multicolumn{2}{l}{Fitted [0.0, 2.0, $10^{-2}$]} \\\hline \hline
\multicolumn{4}{l}{{\bf References:} (1) \cite{2016A&A...595A...2G}, (2) \cite{2006ApJ...653..657M}, (3) \cite{2017A&A...600A..72F},}\\
\multicolumn{4}{l}{(4) \cite{2006AJ....131.2290R}, (5) \cite{2014ApJ...791L..36O}}\\
\multicolumn{4}{l}{$^{*}$ For fitted values, the parameter space  and its resolution (last entry) are stated in brackets.}\\
\multicolumn{4}{l}{$^{**}$  For $\zeta=\infty$, the disk is vertically filled only with large grains.}

\end{tabular}
\end{center}
\end{table*}


\section{Hydrodynamic simulations} \label{hd}
Our symmetrical radiative transfer model cannot explain some features found in the observations, namely the asymmetrical dust distribution, the offsets relative to the central star, and the sub-structures in the bright inner ring, we apply hydrodynamical simulations. Here, we do not aim at precisely fitting HD$\,$169142 but instead at testing the occurrence of these features in a similar but generic protoplanetary disc. For this, we follow the evolution of a non-gravitating viscous protoplanetary disc by means of the public two-dimensional hydrodynamic code 
 \textsc{FARGO-adsg} \citep{2002apa..book.....F, 2008ApJ...678..483B}. It is  a staggered mesh code dedicated to model planet-disc interaction problems, by solving the Navier-Stokes, continuity, and energy equations on a polar grid. We use cylindrical coordinates ($r$,$\phi$) in an equally spaced grid with resolution $r \times \phi = 128 \times 384$. Simulations at higher resolution showed the same results. We modified the \textsc{FARGO-adsg} code in order to include new physical mechanisms such as radiative cooling and stellar heating in the energy equation, using the same prescription as in \cite{2015ApJ...806..253M}.

\subsection{Disc, star, and planets setup}

We consider a generic viscous disc orbiting a 2$\,M_\odot$ mass star with an effective stellar temperature of  $8100\,$K as in our RT model (Section~\ref{rt}). 
For simplicity, we use a low constant viscosity model given by $\nu = 4.5 \times 10^{10} \rm m^2 s^{-1}$, along with a constant opacity prescription $\kappa = 1 ~ \rm cm^2 g^{-1}$.

The disc extends from 4 to 126 AU. The initial surface density profile is given by

\begin{equation}
\Sigma(r) =  7.6 ~ \rm{g ~cm^{-2}} \times  \left( \frac{r}{100\,\rm{au}}  \right) \exp\left(- r / 100\,\rm{au}\right). 
\end{equation}

In order to test whether planets can create the observed structures in HD$\,$169142, we have to start with an unperturbed (gap-less) disk as in opposite to our parametric model (Section~\ref{rt}). The choice for the gap-less density field was taken from \cite{2017A&A...600A..72F}.  In order to find a planet setup which reproduces the ring structure and gaps observed in HD$\,$169142, we applied several combinations of planet masses, $(0.1,0.5,1,2,10)\,$M$_{\rm J}$, and distances from the central star, $(6,15,30,35,45)\,$au, as well as the number of planets ($2-3$). A big 10 $M_{\rm{J}}$ mass planet located at $15\,$au to carve the inner gap, and two $M_{\rm{J}}$ mass planets located at $35\,$au, and $45\,$au are needed to dig the outer large gap. The inner boundary condition is \textit{open} in order to allow gas material to leave the inner disc avoiding gas accumulation. A more quantitative study of the hydrodynamical model is beyond the scope of this study and will be published separately.

\subsection{Post-processing of the hydrodynamic simulation}

\begin{figure}
\centering
\includegraphics[width=\linewidth]{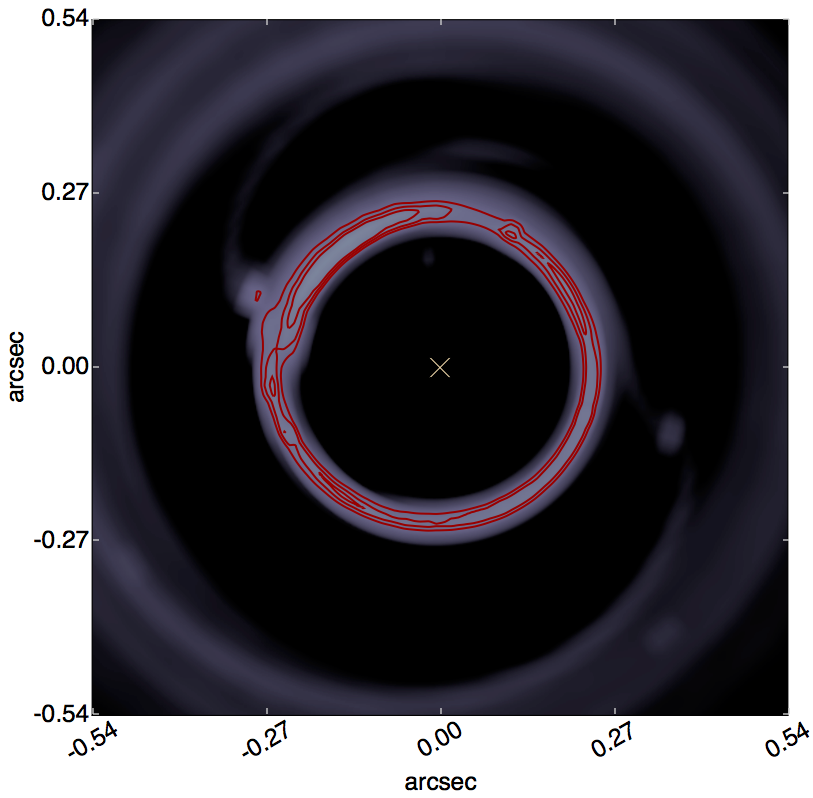}
\caption{Simulated observation at $0.7\mu$m of our hydrodynamical simulation. The contour lines mark the sub-structures in the inner ring which are created by planet-disc interaction. The cross marks the position of the central star.}
\label{pic:hd}
\end{figure}

Using the hydrodynamical simulation data, we compute scattered light images at $0.7\,\mu$m to be compared with observations (Figure~\ref{pic:hd}). For that, we  firstly compute dust opacities for anisotropic scattering light using Mie theory \citep{1983asls.book.....B}, by consensusering spherical dust particles with a dust distribution consistent of a mix of amorphous carbon, and astronomical silicates. Its grain size distribution
follows a power law  $dN/da \propto a^{-3.5}$ \citep{1969JGR....74.2531D, 1977ApJ...217..425M}, with a dust grain population ranging from
$0.1\,\rm{\mu m}$ to 1 cm. We also assume a dust-to-gas ratio 1:100, with the dust  perfectly coupled to the gas, and
the dust temperature equal to the midplane gas temperature i.e., $T_{\rm dust} = T_{\rm gas}$. To extend the two-dimensional surface density of each species to a 3D volume, we assume hydrostatic equilibrium, 
where the disk pressure scale-height is obtained from $H/r = C_{\rm s} / v_{\rm kep}$, where $C_{\rm s}$, and $v_{\rm kep}$ are the sound speed and keplerian velocity respectively.
Finally, we run \textsc{RADMC-3D}\footnote{\url{http://www.ita.uni-heidelberg.de/~dullemond/software/radmc-3d/}} simulations at a wavelength of $0.7\,\mu$m in order to model the emission of small size particles supposed to be coupled to the gas.

\section{Discussions} \label{discussion}

Many features in this disc have been described at several wavelengths in earlier studies. We confirm the findings of a close to face-on inclination as well as a wide gap which separates the disc into a bright inner ring and an outer disc. Further, an additional gap-like structure in the outer disc, which is traced in our observation just at the edge to the $3\,\sigma$~level, coincides with the recently detected gap at $7\,$mm and $9\,$mm, associated with the CO~snowline \citep{macias2017} as well as with coronographic SPHERE/IRDIS observations which are sensitive to the outer disk of HD$\,$169142 \citep{2017ApJ...850...52P}. Our SPHERE/ZIMPOL observation are especially sensitive to the inner disk, and in the following, we discuss our findings for the bright inner ring and the gaps separately.

\subsection{Bright inner ring}
The most intriguing feature in this observation is clearly the bright inner ring. This disc part is not only of striking ring-symmetry in contrast to the rest of this inclined disc. It is also clearly structured (Figure~\ref{pic:obs1},~\ref{pic:obs2}) and offset from the stellar position (see Section~\ref{res}). In this section, we discuss these features in more detail.

To investigate the sub-structures in the ring, we took its surface brightness distribution and subtracted its azimuthal average. The residuals that we recover are shown in Figure~\ref{pic:obs2}. We find brightness bumps  around P.A. $\sim 20^{\circ}$, $\sim 180^{\circ}$, $\sim 320^{\circ}$. The last bump matches the bump observed at  $7\,$mm with the VLA \citep{macias2017}. Please note that the VLA observations present a lack of emission in the North and South direction due to the elongation of the beam. Additionally, the bumps observed with ZIMPOL resemble those described in \cite{2018MNRAS.473.1774L} as well as those which occur in our hydrodynamical simulation as a result of planet-disc interaction (Fig~\ref{pic:sims}). Such bumps are associated with regions of higher density, typical for tidal disruptions induced by forming planets in a disc \citep[e.g.,][]{2014prpl.conf..667B}.  Further, we find that the ring is offset with respect to the central star (see Section~\ref{res}). This offset corresponds to the upper limit of the offset found in the  $7\,$mm data  \citep{macias2017}. In combination with the ring-like structure of the inner disc part, which stands in contradiction to the elliptical appearance of the outer disc in this inclined system, these offsets can be explained by a planet shifting the center of mass outside of the star itself \citep[e.g.,][]{2014prpl.conf..667B}.  This is further supported by our HD~simulations were the chosen setup generates an offset of $0\farcs01125$ from the central star which fits well to the offset derived from the observation (see Section~\ref{res}).
In addition to the offset, examining the radial surface brightness profile (Figure~\ref{pic:polarplot} {\it (d, e)}), we find that the radial distance of the azimuthal brightness peak minimises at two positions ($\sim 45^{\circ}$ and $\sim 190^{\circ}$). This is an interesting feature since a ring of dust, offset from the star, would result in a radial surface brightness profile with only one dip. However, this double-dip structure is characteristic for the case of an intrinsically elliptical dust distribution as it is expected from a scenario in which the central star is orbited by a companion \citep[e.g.,][]{2016ApJ...832...22S, 2017MNRAS.464.1449R}. \cite{2017A&A...600A..72F} propose a giant planet of $0.1-1\,M_{\rm J}$ located inside the bright inner ring based on the lack of azimuthal asymmetries and the drop of the gas surface density in their ALMA data. In this work, however, we do find clear azimuthal asymmetries in the bright inner ring and our HD~simulations support a planet of $10\,M_{\rm J}$ just in the inner edge of the bright ring.

\subsection{Gaps}
We report and confirm two gaps visible in the scattered light images. The innermost gap, however, which appears to be the inner cavity of HD$\,$169142, was identified by modelling the SED and the necessity of a residual dust reservoir close to the star \citep{2012ApJ...752..143H, 2013A&A...555A..64M, 2014ApJ...791L..36O} and can be confirmed by our model, as well (see Section~\ref{rt}).

The gap that separates the disc in a bright inner ring and an outer disc extends from  $\sim0\farcs25$ to $\sim0\farcs47$ ($\sim 29 - 55\,$au), has been described for the first time as annular by \cite{2013ApJ...766L...2Q}. Our observations reveal the elliptical shape of this gap.  With a PA of $5^{\circ}$ its shape fits well to an annular gap inclined by $13^{\circ}$ as found for HD$\,$169142 \citep{2006AJ....131.2290R}. It seems that a scattered signal is created from inside the gap. In our disc model (Figure~\ref{pic:sims}) the gap is completely emptied from dust. Hence, the residual signal in the observation can be explained as an artefact of the convolution with the PSF of SPHERE/ZIMPOL. This is further supported by recent ALMA observations which do not detect any signal from dust grains in this gap \citep{2017A&A...600A..72F}. However, the same ALMA observations show that the gap is filled with gas. The origin of this gap has been discussed in several  studies before. The most likely physical mechanism behind a gap of this size relative to the extent of the disc seems to be planets orbiting the central star and, in doing so, inducing a trapping of radially drifting dust grains at local pressure maxima, depleting the disc along their orbit \citep {1977MNRAS.180...57W, 2014ApJ...780..153B}. \cite{2017A&A...600A..72F} propose a giant planet of $1-10\,M_{\rm J}$ located inside this gap. In our HD~simulations, however, we find that such a planet is not able to open a gap of this extent. Yet, a setup of two planets of $1\,M_{\rm J}$ each results in an appropriate gap size (section~\ref{hd}). Another potential mechanism is the magneto-rotational instability \citep[MRI;][]{1991ApJ...376..214B, 1996ApJ...467...76B, 1998RvMP...70....1B}. MRIs induce low-ionization regions, so-called dead-zones \citep[e.g.,][]{1994ApJ...421..163B, 2012ApJ...761...95F}. Close to the outer edges of these dead-zones large gaps and bump structures can open in the surface density \citep{2015A&A...574A..68F}. However, gaps formed by MRI do typically follow the symmetry of the disc what stands in contradiction to the detected offset of the outer disc relative to the central star (see Section~\ref{res}).

The third gap in HD$\,$169142, is shallow and narrow at $\sim 0\farcs70 (\sim 82\,$au$)$. It coincides with a gap reported by \cite{macias2017} at $7\,$mm and $9\,$mm observed at $\sim 0\farcs73$. We report this gap for the first time at this wavelength. As  \cite{macias2017} discuss, it is likely that this gap is created by an accumulation of dust grains at the position of the CO~snowline.

\section{Conclusion}

We present PDI observations on unprecedented spatial resolution of the circumstellar disc around the Herbig Ae/Be star HD$\,$169142 obtained with SPHERE/ZIMPOL. Our main results are:

\begin{enumerate}
\item The bright inner ring is spatially resolved, for the first time, in the scattered light at short wavelengths. Our observations reveal its complex, irregular sub-structure and its offset position relative to the central star. At the same time, its over-all shape appears widely ring-like shaped which stands in strong contrast to the determined inclination and the rest of this system. Further, we find model-independent evidence that its intrinsic shape is indeed elliptical and we observe it inclined by $13^{\circ}$. 

\item We report a surface brightness dip in the bright inner ring. If this is the same dip which has been observed by  \cite{2013ApJ...766L...2Q}, then its orbital velocity is consistent with a shadow casted by a giant planet or brown dwarf which is surrounded by a small accretion disk and located at $\sim 6\,$au from the central star. Such an object would explain the formation of the inner cavity in HD$\,$169142 and contribute to the residual dust close to the central star which has not yet been spatially resolved.

\item The gap which separates the bright inner ring from the outer disc is of elliptical shape (and not annular, as it was described before in \cite{2013ApJ...766L...2Q}). It can be explained by a circular outer disc which is observed under an inclination of $13^{\circ}$. We find that also the inner rim of the outer disc is offset relative to the central star. Our radiative transfer model shows that the residual scattered light detected inside the gap is indeed a consequence of the convolution with the PSF. Our model supports a gap emptied from dust, as it is supported by ALMA observations.

\item Our radiative transfer model describes a flat, self-shadowed disc which fits the spatially resolved observations obtained with SPHERE/ZIMPOL and ALMA, as well as the SED.

\item Our hydrodynamic simulations show that three giant planets, located at $15\,$au, $35\,$au, and $45\,$au from the central star, can form the bright inner ring with its sub-structures as well as the surrounding gap. Further, we show that this setup can also explain the offsets relatively to the central star found for the inner ring and the outer disc.

\item Our findings indicate ongoing planet-disc interaction in the young Herbig Ae/Be system HD$\,$169142.

\end{enumerate} 

\section*{Acknowledgements}

GHMB, HA, SC, MM, SP, and LC acknowledge support by the Millennium Science Initiative (Chilean  Ministry  of  Economy),  through grant nucleus RC13007. GHMB further acknowledges financial support from CONICYT through FONDECYT grant 3170657, and LC through FONDECYT grant 1171246.



\bibliographystyle{mnras}
\bibliography{lit} 



\bsp	
\label{lastpage}
\end{document}